# Two-photon quantum interference and entanglement at 2.1 µm

Shashi Prabhakar[1], Taylor Shields[1], Adetunmise C. Dada[1], Mehdi Ebrahim[1], Gregor G. Taylor[1], Dmitry Morozov[1], Kleanthis Erotokritou[1], Shigehito Miki[2,3], Masahiro Yabuno[2], Hirotaka Terai[2], Corin Gawith[4,5], Michael Kues[6,7], Lucia Caspani[8], Robert H. Hadfield[1], Matteo Clerici[1]*

Quantum-enhanced optical systems operating within the 2- to 2.5-µm spectral region have the potential to revolutionize emerging applications in communications, sensing, and metrology. However, to date, sources of entangled photons have been realized mainly in the near-infrared 700- to 1550-nm spectral window. Here, using custom-designed lithium niobate crystals for spontaneous parametric down-conversion and tailored superconducting nanowire single-photon detectors, we demonstrate two-photon interference and polarization-entangled photon pairs at 2090 nm. These results open the 2- to 2.5-µm mid-infrared window for the development of optical quantum technologies such as quantum key distribution in next-generation mid-infrared fiber communication systems and future Earth-to-satellite communications.

## INTRODUCTION

The ability to generate and detect quantum states of light underpins the development of secure communications, both for guided wave and free-space systems (1, 2). Free-space quantum key distribution (QKD) has recently enabled quantum-secured intercontinental communication between locations as far as 7600 km apart on Earth (3). Until now, satellite-based QKD has only been possible at night to avoid the blinding effect of the solar background radiation, but recent studies have shown that operating near 1.5 µm could—due to the reduction in solar background radiation (4)—alleviate this issue and enable daylight quantum communications. Moreover, lower scattering losses from subwavelength-sized particles (Rayleigh scattering) further support the use of infrared radiation for free-space communications. At 2.090 µm, the target of this work, the solar irradiance per unit area outside the atmosphere and in a 50-nm bandwidth, is three times lower than the same quantity at 1.5 µm (5), making a compelling case for investigating sources and detectors of nonclassical light fields at longer wavelengths (6).

A further reason to investigate entangled photons at ~2 µm comes from the limitations of guided wave optics (7, 8). The current fiber network faces a "capacity crunch" (9), unless a step change in technology is achieved. Solutions such as novel hollow-core photonic bandgap fibers working at 2 µm with reduced optical nonlinearities and lower losses (10–13) are currently under test for network implementation. Furthermore, integrated photonics is also seeking to expand into the 2-µm region due to the reduced linear and nonlinear losses that are expected for the well-established silicon platform (14, 15). For example, silicon-germanium waveguides have been recently shown to enable communication speeds up to 10 gigabytes/s (16) over a propagation length of nearly a centimeter. In addition, as is the case for free-space communications, the guided wave infrastructure is also bound to adopt a layer of security, and thus, QKD at 2 µm will be required for these new integrated technologies (15).

An additional reason to investigate the generation and detection of nonclassical states of light at 2 µm comes both from sensing (17) and from the emerging field of high-sensitivity metrology and specifically for gravitational wave detection. To extend the reach of the Laser Interferometer Gravitational-Wave Observatory (LIGO) interferometer beyond our galaxy, the LIGO-Voyager upgrade sets an operating wavelength of ~2 µm (18). This is due to the $\lambda^{-2}$ dependence of the scattering from crystalline silicon, which is to be used as a new cryogenically cooled test mass material, and also due to the improvement of optical coatings with amorphous silicon at these wavelengths (19). To improve sensitivity, nonclassical squeezed states of light will be injected in the interferometer. The first step toward a 2-µm "squeezer" for LIGO Voyager was recently shown (18).

## RESULTS

In this work, we show the first generation and characterization of indistinguishable photon pairs and polarization entanglement at 2.1 µm. Degenerate photon pairs are produced via spontaneous parametric down-conversion (SPDC) in a second-order nonlinear crystal. One photon from the pump laser is converted through vacuum noise amplification into two correlated photons at half the energy of the excitation field. To drive this nonlinear process, we used a turn-key compact ytterbium-doped ultrashort-pulse fiber laser (Chromacity Ltd.), emitting a train of pulses with an average power of up to 2.5 W, a repetition rate of 80 MHz, a pulse duration of ~127 fs, and a carrier wavelength of ~1.045 µm. The first test was performed to quantify the generation efficiency of SPDC in a type 0 phase-matched configuration. In this process, the polarization of the excitation field and the generated daughter photons are the same. To this end, we used a 1-mm-long periodically poled, magnesium-doped lithium niobate crystal (MgO-PPLN; Covesion Ltd.). The crystal length was chosen to guarantee maximum conversion efficiency and minimal temporal separation between the pump pulse and the generated SPDC field. The crystal has been poled with the ferroelectric domains periodically inverted to assure coherence between the pump field and the generated photon-pair phase via quasi-phase matching (20) for the whole length of the crystal and over a broad bandwidth. Different poling periods were tested to determine the optimal condition, and the experiments reported here were obtained using a poling period of 30.8 μm and a stable temperature of (30 ± 0.1)°C. We first characterized the source in the high-photon flux regime, using an amplified InGaAs photodiode and phase-locked detection to determine the SPDC efficiency and losses in the system (see Materials and Methods). A 50-nm bandpass filter was used to select the portion of the down-converted field in the degenerate 2090-nm spectral region. Additional filters, such as long-pass (~1.85-μm edge) antireflection-coated germanium windows, were used to reject the intense laser excitation field. The conversion efficiency increases nonlinearly with increasing excitation power, as shown in Fig. 1. The fit to the data follows the standard equation expressing the down-converted power ($P_{SPDC}$) as a function of the pump power, in the phase-matched case, $P_{SPDC} = \alpha \sinh^2(\gamma \sqrt{P_P})$ (21), where $\alpha$ is the loss coefficient, $\gamma$ is proportional to the product of the nonlinear coefficient and crystal length, and $P_P$ is the pump power. This way, we can extract the efficiency $\eta = P_{SPDC}/P_P = (3.0 \pm 0.2) \, 10^{-11}$ at a photon generation rate of ≈0.04 ≪ 1 pair per laser pulse (photon-starved regime) for 10 mW of excitation power. The efficiency at low power is $\eta \simeq \alpha \gamma^2 + \frac{1}{3} \alpha \gamma^4 P_P$, resulting in the expected constant efficiency of SPDC (first term) in this photon-pair generation regime. Such a low rate is essential to avoid the generation of multiple photon pairs per laser pulse, which pose a risk for quantum-secured communication protocols and saturation of the single-photon detectors used to measure the properties of the SPDC radiation. Notably, the driving laser power is enough to generate a bright squeezed vacuum state with >10 photons per pulse (Fig. 1B). We measured a photon purity of 1.8 ± 0.1 modes with a 50-nm bandwidth filter. Single-mode selection is achieved with a 10-nm bandpass filter (see "Characterization of MgO-PPLN crystal efficiency" section in Materials and Methods).

[1]James Watt School of Engineering, University of Glasgow, Glasgow G12 8QQ, UK. [2]Advanced ICT Research Institute, National Institute of Information and Communications Technology, 588-2 Iwaoka, Nishi-ku, Kobe, Hyogo 651-2492, Japan. [3]Graduate School of Engineering Faculty of Engineering, Kobe University, 1-1 Rokkodai-cho, Nada-ku, Kobe-city, Hyogo 657-0013, Japan. [4]Covesion Ltd., Unit A7, The Premier Centre, Premier Way, Romsey, Hampshire SO51 9DG, UK. [5]Optoelectronics Research Centre, University of Southampton, Southampton SO17 1BJ, UK. [6]Hannover Center for Optical Technologies (HOT), Leibniz University Hannover, Hannover, Germany. [7]Cluster of Excellence PhoenixD (Photonics, Optics, and Engineering–Innovation Across Disciplines), Hannover, Germany. [8]Institute of Photonics, Department of Physics, University of Strathclyde, Glasgow G1 1RD, UK.
*Corresponding author. Email: matteo.clerici@glasgow.ac.uk

A crucial step in characterizing the nonclassical properties of the generated SPDC radiation is to demonstrate the generation of correlated photon pairs. The SPDC field is composed of a signal and an idler photon, which are generated together in the nonlinear crystal. They can, however, be separated upon propagation in the far field. Transverse momentum conservation implies that signal and idler photons emerge at opposite angles from the nonlinear crystal, around the excitation-field propagation direction. Signal and idler photons were therefore spatially separated in the far field (in the focus of a parabolic mirror) using a pickoff (D-shaped) mirror, then filtered in bandwidth (50-nm passband), and coupled into a pair of 15-m-long single-mode fibers. A sketch of the experimental setup is shown in the inset in Fig. 2A. The focal length ($f = 50$ mm) of the lens used to pump the nonlinear crystal was chosen to maximize the coupling of signal/idler photons into the single-mode fibers used for the photon-counting experiments, as described in (22). In our case, the Rayleigh length was ≈0.5 mm, that is, half the crystal length $L_c = 1$ mm. The fibers were routed into a cryostat, where superconducting nanowire single-photon detectors (SNSPDs) (23) were used to detect the single photons and time-tagging electronics (PicoQuant HydraHarp) was used to record their arrival time and build the coincidence histogram. The losses of the signal and idler channels upon passing through the filters (within the 50-nm selected window), as well as injection and propagation through the single-mode fiber, amount to nearly −14 dB (see "Characterization of losses" section in Materials and Methods). The twin SNSPD devices each consisted of a backside-illuminated NbTiN nanowire, 6 nm thick and 60 nm wide. These nanowires were meandered over a 15 μm by 15 μm active area and embedded into an optical cavity (24, 25) with dielectric thickness tailored to enhance photon absorption in the mid-infrared. The SNSPD devices were coupled with SM2000 single-mode silica fiber and mounted in a closed-cycle Gifford McMahon cryostat with an operating temperature of 2.5 K. The small SNSPD nanowire cross section, the modified optical cavity design, and SM2000 silica fiber were designed to give improved efficiency at longer wavelengths where conventional single-photon counting technologies are unusable (26, 27). The detection efficiencies of the two SNSPDs at 2.090 μm, including the losses due to fiber coupling into the cryostat, were measured to be $\eta_{det1} = (2.0 \pm 0.2)\%$ and $\eta_{det2} = (1.0 \pm 0.1)\%$. An example of a coincidence histogram (i.e., a histogram showing the arrival time difference of the signal/idler photon on the detectors) for an integration time of 30 min and an input pump power of 41 mW, where the main and accidental coincidence peaks are visible, is shown in Fig. 2A. The histogram shows that the coincidences are maximal at zero delay between the signal and the idler path and that accidental coincidences occur at successive delays of 12.5 ns, that is, the inverse of the laser pulse repetition rate. We note that the pulses are too short to be able to resolve the temporal coincidence peak shape with the timing electronics and detector bandwidth.

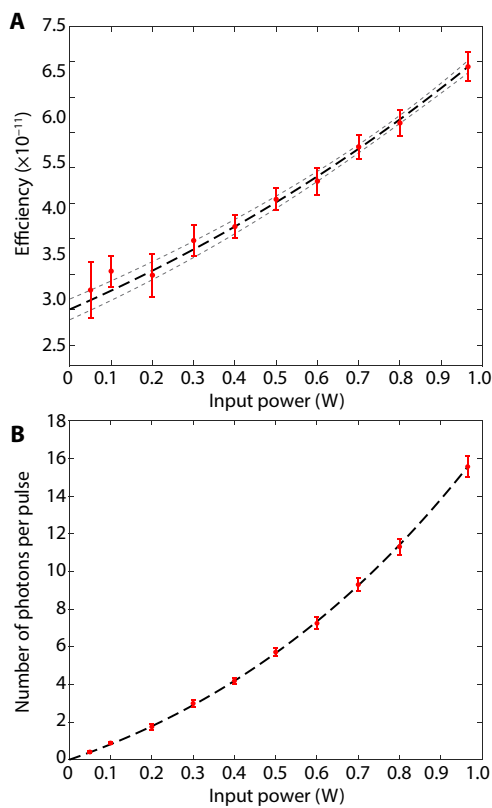

**Fig. 1. Down-conversion efficiency.** Characterization of the spontaneous down-conversion efficiency in a 50-nm bandwidth centered at the 2090-nm degenerate emission wavelength. (**A**) The estimated conversion efficiency in the linear regime is of the order of $(3.0 \pm 0.2) \times 10^{-11}$ (input power of 10 mW). The conversion efficiency is not constant and grows with the input power as expected in a nonlinear fashion (see the main text). The black dashed line is a fit of the conversion efficiency based on the simple model mentioned in the main text. The light gray dashed curves indicate the 95% prediction bounds. In (**B**), we show the measured number of photons generated for input powers up to 1 W. The dashed line is based on the fit in (A).

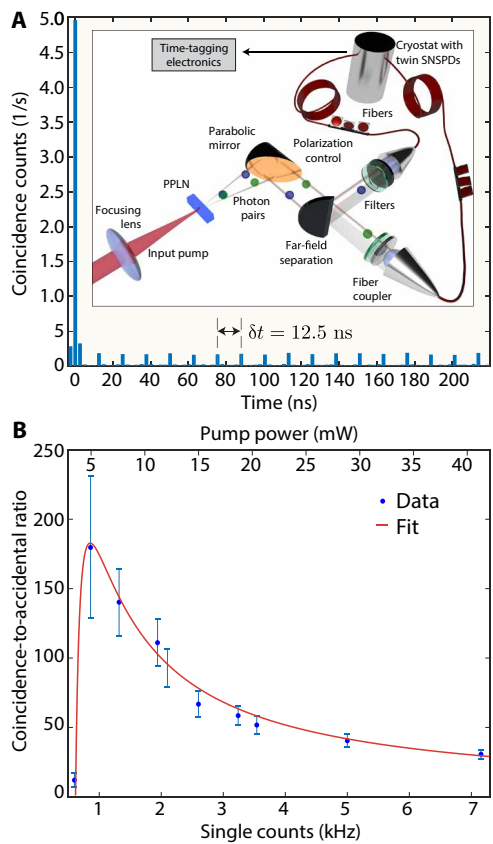

**Fig. 2. Coincidence measurements at 2.1 μm.** (**A**) Coincidence measurement showing the expected peak at zero delay, with accidental peaks at the inverse of the laser repetition rate (~12.5 ns). The bin size considered is ~2.6 ns. The inset shows the experimental setup for this measurement. (**B**) Measured coincidence-to-accidental ratio (CAR) as a function of the averaged single count rates between detectors 1 and 2. The red curve is a fitted CAR based on the model detailed in the main text.

An important figure of merit for correlated photon-pair sources is the coincidence-to-accidental ratio (CAR). In the case of pulsed sources, the CAR is defined as the ratio between the coincidences in the zero-delay peak (signal and idler generated by the same pump pulse) and the average of the coincidences that are recorded between photon pairs from two different SPDC processes, that is, from different pulses. We measured a maximum CAR of $180 \pm 50$ at $\simeq 5$ mW of input pump power (single count rates, $S_1 \sim 960$ Hz and $S_2 \sim 760$ Hz, including dark counts) for 30-min integration time, without background subtraction (integrating over a $\simeq 2.6$-ns window).

The CAR decreases with increasing pump power (see Fig. 2B) as a consequence of the increasing probability of generating multiple photon pairs per pulse, which, in turn, leads to increased accidental counts. At very low input powers, the single and coincidence counts are limited by the detector noise and dark counts (28). As expected, the CAR reaches a maximum value and then decreases as $1/P_p$ (29) and can be modeled as CAR = $\mu\,\eta_1\eta_2/(\mu\,\eta_1 + d_1)(\mu\,\eta_2 + d_2) + 1$, where $d_1$ and $d_2$ are the dark count probabilities per pulse for channels 1 and 2, respectively, $\eta_1$ and $\eta_2$ are the total channel efficiencies, and $\mu$ is the average photon-pair number per pulse (30). In our case, $d_1$ and $d_2$ are determined by the dark count rates ($D_1 \simeq 600$ Hz and $D_2 \simeq 550$ Hz), the time bin window and the laser repetition rate $R = 80$ MHz. $\eta_1$ was measured as the product between the filters, the fiber propagation, the fiber coupling, and the detection losses (see "Characterization of losses" section in Materials and Methods). The two detector efficiencies were measured, and the filters and fiber propagation losses are expected to be the same for channels 1 and 2. The model of the CAR described above suggests a mismatch between the measured and expected losses. To quantify this difference, we used the coupling losses of channel 2 as a free parameter for the fit. The result shown by the red curve in Fig. 2B is obtained for a fiber coupling efficiency of $\simeq 0.31$, nearly three times higher than what was measured. This is consistent with the experimental inaccuracies, as the coupling is highly sensitive to fine alignment of the setup and can vary from experiment to experiment.

To determine the quality of indistinguishability of the generated photons, we investigated the two-photon interference using a Hong-Ou-Mandel (HOM) interferometer (31). In such a measurement, signal and idler photons are injected into the two input ports of a 50:50 beam splitter, respectively, and the photon coincidences are measured between the output ports. If the photon pairs are indistinguishable in all the degrees of freedom and their paths overlap, then they will exit together from one of the two output ports and will never end up in different output ports, effectively cancelling the coincidence signal. To perform this measurement, we introduced in the experimental setup shown in Fig. 3A a polarization-maintaining fiber-based 50:50 beam splitter and recorded the coincidences at the beam splitter output when varying the delay of one input with respect to the other (a sketch is shown in Fig. 3A). The characteristic HOM dip is shown in Fig. 3B. The asymmetry visible in the regions outside of the HOM dip is most likely due to a variation in coincidence counts over the length of the entire set of measurements. The raw experimental data are fitted with an inverted Sinc function weighted with a Gaussian (see "HOM dip" section in Materials and Methods). We achieved a visibility of 88.1%, demonstrating photon indistinguishability via two-photon interference, which lies at the core of photonic quantum information processing.

Polarization-entangled photons are one of the pillars of quantum technologies, e.g., for QKD. To realize a polarization entangled state, we generated signal and idler photons with orthogonal polarizations. To this end, we fabricated a 300-μm-long MgO-PPLN crystal cut to perform type 2 (e-oe) down-conversion, with a poling period of 13.4 μm for phase matching at 110 °C. The type 2 crystal was shorter than the type 0 one to minimize the impact of the group velocity mismatch between the orthogonally polarized signal and idler on the measurements. Such a system would generate a polarization-encoded Bell state of the form

$$| \psi^- \rangle = \frac{1}{\sqrt{2}}\left(| H,V\rangle - | V,H\rangle\right)$$

We show entanglement using a test of the violation of the Clauser-Horne-Shimony-Holt (CHSH)–Bell inequality $S \leq 2$ (32), which involves polarization-correlation measurements between the signal and idler photons. Within the framework of quantum mechanics, a violation of this inequality will be a sufficient proof that the state is entangled and shows that the correlations of the signal and idler photons persist irrespective of the measurement basis states. The experimental setup for demonstrating entanglement is illustrated in Fig. 4A. We use vertical polarizers preceded by half-wave plates oriented at angles $\theta_a/2$ and $\theta_b/2$ in each of the signal and idler paths, respectively. This allows us to project the signal and idler photons to the measurement basis states $|\theta_a\rangle$ and $|\theta_b\rangle$, respectively. For a maximally entangled signal-idler state $|\psi^-_{\mathrm{ME}}\rangle$, the coincidence count rate is given by $C(\theta_a, \theta_b) \propto |\langle \psi^-|\theta_a, \theta_b\rangle|^2 = \sin^2(\theta_a - \theta_b)$. The CHSH-Bell parameter can be written as (33)

$$S = E(\theta_a, \theta_b) - E(\theta_a, \theta'_b) + E(\theta'_a, \theta_b) + E(\theta'_a, \theta'_b)$$

where the correlation coefficients are defined as

$$E(\theta_a, \theta_b) =$$
$$\frac{C(\theta_a, \theta_b) + C\left(\theta_a + \frac{\pi}{2}, \theta_b + \frac{\pi}{2}\right) - C\left(\theta_a + \frac{\pi}{2}, \theta_b\right) - C\left(\theta_a, \theta_b + \frac{\pi}{2}\right)}{C(\theta_a, \theta_b) + C\left(\theta_a + \frac{\pi}{2}, \theta_b + \frac{\pi}{2}\right) + C\left(\theta_a + \frac{\pi}{2}, \theta_b\right) + C\left(\theta_a, \theta_b + \frac{\pi}{2}\right)}$$

and, similarly, for $E(\theta_a, \theta'_b)$, etc. We measured coincidences for the 16 combinations of orientation angles (see Fig. 4B), which would result in maximal violation of the inequality ($S = 2\sqrt{2}$) in the case of a maximally entangled singlet state $|\psi^-\rangle = \frac{1}{\sqrt{2}}(|H, V\rangle - |V, H\rangle)$, i.e., using $\theta_a = 0$, $\theta_b = \frac{\pi}{8}$, $\theta'_a = \frac{\pi}{4}$, and $\theta'_b = \frac{3\pi}{4}$. For our source, we determined the Bell parameter to be $S = 2.20 \pm 0.09 > 2$, clearly demonstrating entanglement. The value of the measured $S$ parameter corresponds to a visibility of $V = 77.8 \pm 0.1\%$ for the coincidence count rate $C(\theta_a - \theta_b)$, where $V = (C_{\max} - C_{\min})/(C_{\max} + C_{\min})$. On the basis of the channel losses and an average detection rate of $3107 \pm 55$ Hz in each channel, we estimate the entangled photon generation rate to be ~200 kHz at the source. However, the coincidence count rate within the 512-ps time window used for the obtained Bell inequality violation is $0.20 \pm 0.01$ Hz, mainly due to photon losses. The rate of entangled photons can potentially be increased by an order of magnitude by improving/increasing the detection efficiencies at 2.1 µm and reducing the in-line losses. Doing so will also increase the visibility, which is currently affected by spurious coincidences due to detector dark counts (≈550 Hz).

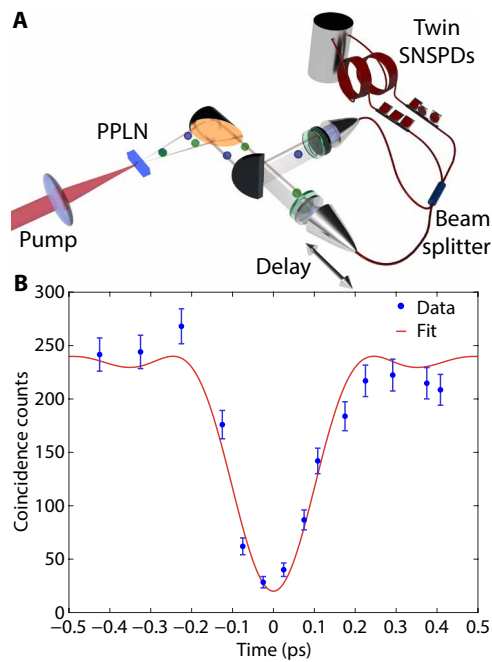

**Fig. 3. Two-photon interference.** (**A**) Experimental setup for characterizing the two-photon interference. A beam splitter is inserted in front of the coincidence detection together with a tunable delay line allowing the adjustment of the temporal overlap of the down-converted photons at the beam splitter. (**B**) The observed two-photon interference (HOM dip) where the dots represent the experimental twofold coincidence counts and the solid curve was the fit to the experimental data (see Materials and Methods for details). Errors were estimated assuming Poisson statistics.

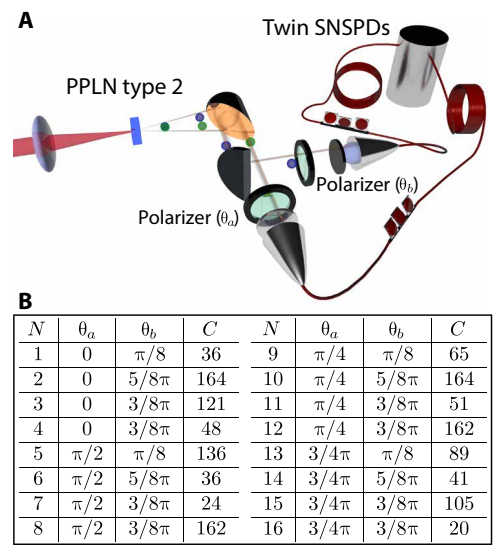

**Fig. 4. Polarization entanglement.** (**A**) Experimental setup for entanglement characterization. A tunable polarizer is inserted before each single-photon detection, and coincidences measured for different settings (angles), allowing the demonstration of a violation of the CHSH-Bell inequality (see the main text for details). (**B**) Measurement settings for the CHSH-Bell test and raw coincidence counts ($C$) used to determine the Bell parameter $S = 2.20 \pm 0.09$, demonstrating genuine polarization entanglement at 2.1 µm (see the main text for details). The integration time was 30 min for each measurement.

| $N$ | $\theta_a$ | $\theta_b$ | $C$ | $N$ | $\theta_a$ | $\theta_b$ | $C$ |
|---|---|---|---|---|---|---|---|
| 1 | 0 | $\pi/8$ | 36 | 9 | $\pi/4$ | $\pi/8$ | 65 |
| 2 | 0 | $5/8\pi$ | 164 | 10 | $\pi/4$ | $5/8\pi$ | 164 |
| 3 | 0 | $3/8\pi$ | 121 | 11 | $\pi/4$ | $3/8\pi$ | 51 |
| 4 | 0 | $3/8\pi$ | 48 | 12 | $\pi/4$ | $3/8\pi$ | 162 |
| 5 | $\pi/2$ | $\pi/8$ | 136 | 13 | $3/4\pi$ | $\pi/8$ | 89 |
| 6 | $\pi/2$ | $5/8\pi$ | 36 | 14 | $3/4\pi$ | $5/8\pi$ | 41 |
| 7 | $\pi/2$ | $3/8\pi$ | 24 | 15 | $3/4\pi$ | $3/8\pi$ | 105 |
| 8 | $\pi/2$ | $3/8\pi$ | 162 | 16 | $3/4\pi$ | $3/8\pi$ | 20 |

## DISCUSSION

We have demonstrated an efficient quantum light source in a free space setup configuration at 2.090 µm and verified indistinguishable photon pairs through two-photon interference. The HOM dip with high visibility proves that this source can be successfully used in applications requiring interferometric measurements at ~2 µm. We have demonstrated that polarization-entangled photon pairs can be generated, manipulated, and detected with our approach. This work provides a new platform for quantum optics and paves the way for technological applications for quantum sensing and quantum-secured long-distance communications within this wavelength regime. Advancements in silica-based fiber optics indicate that operating at 2 µm minimizes transmission losses potentially offering a solution to concerns regarding network capacity limits with current fiber-optic infrastructure (11, 34). In addition, since the security of any device-independent QKD protocol necessarily relies on the violation of a Bell inequality (35), this result opens the pathway to technological applications for QKD in the ~2-μm window. Free-space daylight QKD at 2.1 μm could offer an additional advantage of a threefold reduction in blackbody radiation compared with telecom alternatives. More ambitious experimental and system demonstrations will require engineering of improved single-photon components at ~2 μm. Future work shall address improved detection at this wavelength, taking advantage of rapid progress in superconducting photon-counting technologies and schemes whereby the infrared photons are frequency up-converted to a spectral region where single-photon detection efficiencies are already close to unity (36, 37).

## MATERIALS AND METHODS

### Characterization of MgO-PPLN crystal efficiency

We characterized the MgO-PPLN crystal and determined the optimal temperature for maximizing the down-conversion events from the crystal using an InGaAs detector (Thorlabs PDA10DT-EC). To enhance the signal-to-noise ratio, we used an optical chopper and a lock-in amplifier to detect the generated signal. Characterization of the crystal was performed by pumping with the full laser power of 965 mW (measured after the optical chopper) and then varying the crystal temperature. On the basis of this and because of temperature-control limitations, we chose an operating temperature of 30°C. At maximum power, the number of photons generated per pulse was ≃16, indicating that a power of <50 mW would be required to reach the photon-starved regime. The crystal temperature was stabilized within 0.1°C accuracy to optimize the collinear generation of photon pairs at the degeneracy wavelength of 2.090 μm. Similar characterization was performed for the type 2 crystal used for polarization entanglement.

### Characterization of losses

The generation efficiency of the type 0 crystal was measured by lock-in detection, where a value of $3.0 \pm 0.1 \times 10^{-11}$ was obtained. The measured in-band transmission of the antireflection-coated germanium filter, the 50-nm passband filter, and 15 m of SM2000 fiber at a wavelength of 2 μm were $0.808 \pm 0.001$, $0.65 \pm 0.01$, and ≃0.8, respectively. The total coupling efficiency of the SPDC into the SM2000 fiber (after passing the D-shaped mirror) was measured for channel 1 to be $0.10 \pm 0.05$. The overall transmission from the output of the crystal to the input of the cryostat for channel 1 was, therefore, ≃0.04 ± 0.01 (−14 dB). The coupling efficiency for channel 2 was obtained by the fit reported in Fig. 2B and was ≃3.1 times that for channel 1. The overall transmission from the output of the crystal to the input of the cryostat for channel 2 was, therefore, ≃0.12 (−9 dB). The measured detection efficiency of the two SNSPDs, including the losses of the access fibers inside the cryostat, was ≃2.0 ± 0.2% for detector 1 and ≃1.0 ± 0.1% for detector 2.

### Single-photon detectors

The SNSPDs used for the coincidence measurements had dark count rates of ≃600 Hz for channel 1 and ≃550 Hz for channel 2, which contributed to the accidental coincidence peaks, especially in the photon-starved regime. To optimize the absorption in the mid-infrared spectral region, the SNSPDs were fabricated with an optical cavity incorporating a 600-nm-thick $SiO_2$ dielectric layer. The coincidence histogram was acquired with a time-to-digital converter/time-correlated single-photon counting module (PicoQuant HydraHarp 400). We recorded the individual coincidence peaks with 256-ps time resolution (time bin window). During the analysis, 10 bins were combined with an overall bin time of ≃2.6 ns. The SNSPDs were characterized at 2.090 μm using an optical parametric oscillator (Chromacity Ltd.) as a source. The spectrally broad output was directed through a narrow bandpass filter (the same filter as in the main experiment) and then attenuated using nondispersive filters. In this way, the power at the cryostat input port could be carefully calibrated. The closed cycle Gifford McMahon cryostat was operated at 2.5 K, and the detectors were coupled inside with the same type of single-mode fibers used elsewhere in the experiment (SM2000). The input photon flux was calibrated to be 0.5 photons per pulse. Efficiency measurements were calculated by subtracting the dark count rate from the photon count rate and dividing by the incident photon flux. No polarization control was introduced for this measurement. The same quantum efficiency values were obtained by an independent calibration performed directly with the SPDC radiation filtered around degeneracy (50-nm bandwidth). The polarization dependence of the SNSPDs (measured using an optical parametric amplifier source at a wavelength of 2.3 μm) was ±25%. In the Bell test measurements, the input into the SNSPDs was adjusted by fiber polarization control as shown in the figures for maximum coupling to each SNSPD, thereby removing the effect of any polarization dependence of the individual detectors on the measurements.

### HOM dip

The uncorrected experimental data are fitted with an inverted Sinc function weighted with a Gaussian function, and this fit was used to facilitate the estimation of the visibility. The shape of the HOM dip is given by the Fourier transform of the joint spectral amplitude, which, for pulsed SPDC, is given by the product of the pump pulse envelope (Gaussian) and phase-matching function (38). In our case, the phase matching is filtered by a square filter (50-nm bandwidth centered at 2090 nm) resulting in the product of a Gaussian and Sinc function for the HOM dip. This is the profile expected, on the basis of the theory of pulsed SPDC (38–40). The visibility of 88.1% was conservatively determined using the lowest raw experimental data point within the dip as the minimum coincidence value and an asymptotic value of the fitting function as the maximum coincidence value. The main challenge with this realization of the HOM interference (shown in Fig. 3A) was obtaining parity between the small-but-finite difference in the lengths of polarization-maintaining input fiber-based 50:50 beam splitter. These fibers were designed for wavelengths of 2080 ± 40 nm. For finer delay control, the fiber couplers with filters were mounted on motorized translational stages with a minimum incremental motion of 50 nm.

### Photon purity

One relevant aspect of photon-pair sources, especially in the framework of heralded single-photon applications, is the number of modes composing the biphoton field. Only if signal and idler photons are generated in a single mode will the measurement of one project the other in a single-photon state with high purity (41), rather than a mixed state. Since only one spatial mode is guided in the fiber, we measured the number of temporal modes performing a Hanbury-Brown and Twiss measurement on the signal filed [intrabeam $g^{(2)}$]. This is achieved by recording the coincidences at the output ports of the 50:50 beam splitter when only the signal field is injected. Given the thermal statistics of the signal or idler field when measured independently (reminiscent of the amplification of vacuum fluctuations), the peak of the $g^{(2)}$ is related to the number of modes $M$ by the relation [42] $g^{(2)}(0) = 1 + 1/M$. Our measurements show that, without a bandpass filter and only relying on the spectral selectivity of the Germanium filter and of the other components of the setup, the number of modes is $M = 7 \pm 2$. Adding the 50-nm passband filter centered at 2090 nm, which is the condition of the experiments reported in the Results section, the number of modes dropped to $M = 1.8 \pm 0.1$. Last, adding a 10-nm filter centered at the degenerate wavelength, we can ensure single mode: $M = 1.01 \pm 0.06$. We note that the number of modes is approximately given by the ratio between the filter bandwidth and the pump pulse bandwidth.

## Variance on the S parameter

Error bars are estimated on the basis of Poissonian counting statistics. Since the Bell parameter $S$ is a function of 16 coincidence counts $C_j$ with $j = 1, \ldots, 16$ corresponding to all combinations of measurement angles, we obtain the uncertainty in $S$ using

$$\sigma_S = \sqrt{\sum_{j=1}^{16} \left( \sigma_{C_j} \frac{\partial S}{\partial C_j} \right)^2}$$

where $\frac{\partial S}{\partial C_j}$ is the partial derivative of $S$ with respect to $C_j$. The uncertainty in $C_j$ is given by $\sigma_{C_j} = \sqrt{C_j}$, based on Poissonian counting statistics.

**Acknowledgments:** We acknowledge D. Faccio, M. Sorel, A. Casaburi, M. Goossens, C. G. Leburn, and Chromacity Ltd. for enlightening discussions and technical support. **Funding:** M.C., S.P., R.H.H., and C.G. acknowledge the support from Innovate UK (project PEPE EP/R043299/1). M.C. and A.C.D. acknowledge the support from the UK Research and Innovation (UKRI) and the UK Engineering and Physical Sciences

Research Council (EPSRC) (Fellowship "In-Tempo" EP/S001573/1). A.C.D. acknowledges support from the EPSRC, Impact Acceleration Account (EP/R511705/1). C.G. acknowledges financial support from the Royal Academy of Engineering Senior Research Fellowship scheme. M.K. acknowledges support from the BMBF in the framework of the Quantum Futur program. R.H.H. acknowledges support through the European Research Council Consolidator Grant (IRIS 648604), the UK EPSRC (grants EP/L024020/1 and EP/M01326X/1), and the Royal Society Leverhulme Trust Senior Research Fellowship. S.M., M.Y., and H.T. acknowledge the support from MEXT Q-leap program (grant JPMXS0118067634). K.E. acknowledges an internship at NICT Japan. **Author contributions:** S.P., T.S., and M.C. designed the experiment. S.P., T.S., A.C.D., G.G.T., D.M., M.K., M.E., L.C., and M.C. performed the measurements and analyzed the results. K.E., S.M., M.Y., H.T., and R.H.H designed, fabricated, and characterized the SNSPDs. C.G. designed and fabricated the nonlinear PPLN crystal. M.C., S.P., A.C.D., T.S., M.K., L.C., and R.H.H. drafted the manuscript. M.C. supervised the project. All the

# Science Advances

**Two-photon quantum interference and entanglement at 2.1 μm**

Shashi Prabhakar, Taylor Shields, Adetunmise C. Dada, Mehdi Ebrahim, Gregor G. Taylor, Dmitry Morozov, Kleanthis Erotokritou, Shigehito Miki, Masahiro Yabuno, Hirotaka Terai, Corin Gawith, Michael Kues, Lucia Caspani, Robert H. Hadfield and Matteo Clerici

| | |
|---|---|
| **ARTICLE TOOLS** | http://advances.sciencemag.org/content/6/13/eaay5195 |
| **REFERENCES** | This article cites 36 articles, 2 of which you can access for free<br>http://advances.sciencemag.org/content/6/13/eaay5195#BIBL |
| **PERMISSIONS** | http://www.sciencemag.org/help/reprints-and-permissions |